# Observation of a Majorana zero mode in a topologically protected edge channel

Berthold Jäck†[1], Yonglong Xie†[1], Jian Li†[2,3], Sangjun Jeon[1,4]( 전상준), B. Andrei Bernevig[1], Ali Yazdani*[1]

[1] Joseph Henry Laboratories and Department of Physics, Princeton University, Princeton, NJ 08544, USA

[2] Institute for Natural Sciences, Westlake Institute for Advanced Study, Hangzhou, Zhejiang, China.

[3] School of Science, Westlake University, Hangzhou, China.

[4] current address: Department of Physics, Chung-Ang University, Seoul, 06974, Republic of Korea

*yazdani@princeton.edu
† authors contributed equally

**Abstract:** Superconducting proximity pairing in helical edge modes, such as those of topological insulators (TI), is predicted to provide a unique platform for realizing Majorana zero modes (MZMs). We use scanning tunneling microscopy measurements to probe the influence of proximity induced superconductivity and magnetism on the helical hinge states of Bi(111) films, grown on a superconducting Nb substrate and decorated with magnetic Fe clusters. Consistent with model calculations, our measurements reveal the emergence of a localized MZM at the interface between the superconducting helical edge channel and the Fe clusters with strong magnetization component along the edge. Our experiments also resolve the MZM's spin signature that distinguishes it from trivial in-gap states that may accidently occur at zero energy in a superconductor.

**Main Text:** Majorana zero modes (MZMs) are non-Abelian quasi-particles that can emerge at the ends of one-dimensional (1D) topological superconductors (*1,2*). They may one day provide an experimental setting for topologically protected qubits. To date, strong evidence for the existence of MZMs has come from systems in which the proximity effect from a conventional superconductor is used in concert with strong spin-orbit, Zeeman or ferromagnetic exchange interaction to engineer a system with topological superconductivity. In particular, experiments on semiconducting nanowires and magnetic atomic chains have provided key signatures of MZMs (*3-5*) and of their unique charge and spin spectroscopic properties (*6-8*). The presence of MZMs in these platforms however does not rely on intrinsically topological electronic states but rather on engineering 1D band structures with properties suitable for an "effective" p-wave order parameter to emerge when in contact with a conventional s-wave superconductor (*9-14*). The strength of the induced pairing in these systems, which is the key parameter for the protection of a MZM from quasi-particle poisoning, is also reliant on the strength of spin-orbit interaction. As an alternative to these systems, one might consider topological insulators (TIs), the helical edge modes of which intrinsically exhibit large (essentially infinite at low energies in the bulk gap) spin-orbit interaction. Crucially, TIs provide a platform where topological superconductivity can be realized in channels that are protected by time-reversal symmetry, allowing for material imperfections (*15*) in the absence of a magnetic field. Because this concept does not rely on band engineering, and the presence of a magnetic field is not necessary in the superconducting coupled part of the helical edge mode to induce MZMs, the strength of the induced pairing on such channels could also be large. Thus this potential MZM platform could provide better protection than existing schemes from both disorder and thermal excitation of quasi-particles (*16*).

**Hinge states as an MZM platform**

To realize topological superconductivity and MZMs in a time-reversal-protected helical channel, we utilize the topological edge and hinge states of bismuth (Bi). Although it has long been recognized that a Bi bilayer in isolation is a two-dimensional (2D) topological insulator with 1D helical edge modes (*17*), recent efforts have argued that bulk Bi is an example of a higher order topological insulator (*18-22*), which hosts topologically protected 1D helical states along its hinges (*23*). Consistent with this new perspective, scanning tunneling microscopy (STM) studies on the Bi(111) surface have shown that on the perimeter of a hexagonally shaped bilayer island, every other edge shows signatures of 1D topological boundary mode (*24*). This experimental observation can be understood in terms of the hinge modes being hybridized with the surface state of Bi(111) along half of the edges of the hexagonal island (Fig. 1A). Other measurements have also probed Josephson coupling through the Bi hinge states in a nanowire geometry, demonstrating ballistic transport consistent with their topological nature (*25*). Here, we induce superconductivity in these topological hinge states (*26-28*) and terminate them using a ferromagnetic cluster, which breaks local time-reversal symmetry. The superconductor-ferromagnetic-cluster interface represents a domain wall where the gap changes its character and where we expect a MZM to emerge (Fig 1A).

We theoretically consider this MZM platform using a tight-binding model to capture the 1D topological edge states of a bilayer island on a Bi(111) surface and a proximity induced pairing on such edge channels via coupling to a conventional bulk superconductor (*29*); we also account for their hybridization with both the electronic states of a ferromagnetic cluster and those of the underlying Bi bulk, which leads to the hinge-like behavior of these states (see Sec. 1 of (*30*)). The results of our theoretical modeling confirm our expectation for the existence of a MZM in the superconducting topological edge states of Bi (Fig.1). They also show additional features that have not been considered in previous proposals (*15*, *26*) involving topological edge modes, which, however, are important to the experimental implementation. As expected, we find a proximity-induced superconducting gap in the local density of states (LDOS) along the hinge state and a MZM that emerges from this edge channel at the boundary between the regions of dominant induced pairing and magnetic gaps (Fig. 1, A to C). We also find that the hybridization of the hinge states with the Bi bulk states along half of the bilayer edges (Fig. 1A) renders the wavefunction of the second MZM at the interface with the magnetic cluster highly delocalized with weak contributions to the LDOS at such locations (See Sec. 1 of (*30*)). Our calculation also highlights the presence of in-gap Shiba states, which can occur near magnetic impurities in a superconductor and appear on the magnetic cluster in the LDOS at finite energy (Fig. 1, B and C) (*31*, *32*). These trivial states require a finite magnetic polarization regardless of its spatial orientation. In contrast, as discussed below, the appearance of an MZM is in addition sensitive to the orientation of the magnetism of the cluster. This difference as well as distinctions in the spin properties of the MZM from those of Shiba states can be used to distinguish an MZM from a trivial zero energy Shiba state.

Previous proposals considered a local Zeeman interaction for the opening of magnetic gap at the time-reversal invariant points of the topological edge state band structure (*15*). In our platform, the hybridization of the Bi edge channels with the electronic states of a ferromagnetic cluster (such as iron (Fe) in our experiments) results in an additional magnetic hybridization gap $\Delta_H$ in the topological edge state band structure ( Fig. 1D). Thanks to the direct coupling between the cluster and the edge, a magnetic hybridization gap can have a much larger magnitude than a Zeeman-induced magnetic gap, which enters as a second order effect in terms of the same coupling. Therefore, in our platform we can circumvent the need for fine tuning the chemical potential into the magnetic gap for an MZM to emerge. Additional calculations also reveal that the opening of such magnetic hybridization gaps in the proximitized topological band structure is indeed accompanied by a change of the system's topological invariant, as is required for the topological protection of the localized MZM (see Sec. 2 of (*30*)). However, as shown in Fig. 1, D to E, the appearance of such a magnetic hybridization gap by coupling the edge mode to the magnetic cluster not only depends on the strength of the coupling but also on the cluster's magnetic exchange energy scale and the orientation of its magnetization. Although the exact spin polarization of our Bi helical edge mode is not known, similar to other spin-momentum-locked topological edge modes, we expect the spin polarization to point along some direction orthogonal to momentum of these 1D states along the edge. Therefore, a cluster with a magnetization pointing along the edge is most favorable for opening of a gap in the edge state's band structure so as to localize a MZM (Fig. 1D). Additional magnetization components of such clusters modify the band structure in more complex ways and are usually detrimental to opening up a gap at the Fermi level (Fig. 1E). This sensitivity to the cluster's magnetization shows that MZMs can only emerge for clusters with properly oriented magnetization and ultimately provides the potential ability to tune MZMs on and off on this platform, by reorienting the magnetization.

**Realization of the platform**

To realize our proposed platform, we grow thin films of Bi(111) on a (110) oriented single crystal of niobium (Nb) (sputtered and annealed first for cleaning) and use in-situ low temperature STM imaging and spectroscopy at 1.4 K to confirm the presence of topological hinge states on their surface (see Materials and Methods in (*30*)). As shown in Fig. 2, these states can be identified on the surface of 30 Å thick Bi(111) thin films by van Hove singularity in their LDOS, as detected by STM spectroscopy (Fig. 2B) and its mapping (Fig. 2D). They appear on edges we label as type A, where they are not hybridized with the bulk or surface state of Bi crystals. Spectroscopic mapping along finite segments of these edge channels near the Fermi energy shows standing wave patterns in their LDOS (Fig. 2 E), the Fourier transform of which reveals the dispersion of a 1D band near the Fermi energy (see Sec. 3 & 4 of (*30*)). A previous analysis of these standing wave pattern shows that the observed wavelengths are determined by the scattering between non-orthogonal spin states of the edge state band structure within one half of the Brillouin zone (*24*). Scattering and interference between time-reversed pairs of states in the edge band structure is not observed in such measurements—a behavior consistent with the absence of backscattering in such topological states.

Owing to the proximity effect from the underlying Nb substrate, superconductivity opens a gap in the 1D band structure at the Fermi energy that can be distinguished from the gap found on the bilayer surface (Fig. 2E). We can separate the gap measured on the A edge into two contributions coming from the surface state and the topological edge state, of which the latter yields a topological gap that can be fitted using BCS theory (see Sec. 5 of (*30*)). Following this analysis, we find the gap of the topological edge state ($\Delta_T = 1.50 \pm 0.01$ meV) to have approximately the same size (slightly lower, a point which will be later used) as that measured on the bare Nb(110) substrate surface ($\Delta_S = 1.52 \pm 0.01$ meV). Given that such superconducting gaps in topologically protected helical channels are intrinsically non-trivial, this system has, by far, the largest gap magnitude as compared to other 1D MZM proximity platforms, such as atomic chains or semiconductor nanowires reported to date, which have $\Delta_T \leq 200$ μeV *(4,5)*. The application of magnetic field or presence of ferromagnetism reduces the p-wave gap considerably in these systems as compared to that of conventional superconductor used for proximity. One other example of intrinsic topological superconductivity with large gaps (2meV) has been reported in the Fe-based superconductors, where signatures of MZMs within the vortex cores of their 2D topological surface states have been recently observed *(33, 34)*.

To introduce a local magnetic perturbation that results in the emergence of MZMs along the proximitized Bi hinge states, we deposit Fe atoms on the Bi(111) surface and anneal the surface (to 373 K) after which we find clusters of Fe nucleated at the bilayer step edges of our sample. The most common location for the nucleation of such clusters is between A and B edges as shown in Fig. 3. While we find some clusters at the center of a B edge (see Sec. 10 and Fig. S14 & S15 of *(30)*), we never find clusters at the center of an A edge. The clusters appear with different lengths with a height 2-3 Å (as measured from the top bilayer) in the STM topographies. The magnetic properties of such clusters and their influence on the electronic properties of the topological edge state are characterized by means of spin-polarized STM (see Sec. 6 of *(30)*) *(35)*. In such measurements, we record the spin-polarized tunnel conductance d$I$/d$V$ at low bias voltage between the sample and an Fe-coated Cr STM tip as a function of the applied magnetic field. The d$I$/d$V$ measurement on the Fe cluster, hereafter labelled #1, displays a step-like change of the conductance, ΔG, near zero field, which is characteristic of tunneling between a ferromagnetic cluster and a superparamagnetic tip (Fig. 3B). The small hysteric behavior in these conductance measurements confirms the weak magnetic anisotropy of our tip's magnetization and its superparamagnetic behavior. Spin-polarized measurements performed with such tips away from the magnetic cluster along the bilayer A edge also show similar step-like features in d$I$/d$V$. Characterizing these measurements by plotting the relative change of the conductance Δ$G$ with respect to the conductance at small positive magnetic field as a function of the tip distance from the cluster along the A edge (Fig. 3C), we find a behavior indicative of a decaying oscillatory Ruderman-Kittel-Kasuya-Yoshida-interaction-induced polarization of the one-dimensional edge state by the Fe cluster (*36*, *37*). The coupling of the Fe clusters to the Bi electronic states appears to also influence the clusters' LDOS—leaving spectroscopic features that can be captured by our model and used to extract the strength of the magnetic hybridization gap in the Bi-edge modes caused by the Fe clusters (see Sec. 7 of (*30*)). Together, these measurements demonstrate the ferromagnetic nature of the Fe clusters, their ability to induce magnetism into the topological edge state of Bi, and indications of the clusters introducing a sizable magnetic hybridization gap on the Bi-edge modes (~80meV, see Sec. 7 of (*30*)).

Examining different Fe clusters, we observe a similar behavior; however, we also find evidence for variations in the magnetization axis of different clusters. Characterizing STM topographies measured with the superparamagnetic tips that are polarized using our in-situ vector-magnetic field, we can reconstruct the effective magnetization vector of the clusters (see Sec. 8 of (*30*)). Our results for clusters labeled #1-#6 are summarized in Fig. 3D, showing that these clusters can be separated into two groups (orange and blue) by the amplitude of their magnetization component perpendicular to the edge. Such different cluster magnetizations may impact the 1D edge modes band structure in multiple ways, as anticipated by our model calculations shown in Fig.1 (see Sec. 1 of (*30*))

**Observation of MZMs**

Having established the presence of both superconducting pairing and ferromagnetism in our samples, we use high-resolution STM spectroscopy to probe the low energy properties and to search for localized MZMs in this system. In Fig. 4, we show a typical example of such an experiment on cluster #2 that shows the emergence of a sharp zero bias peak (ZBP) within the pairing gap along the Bi bilayer A edge (peak width $\Delta E \leq 400$ µeV, Fig. 1, B-D) as the tip approaches the hinge-cluster interface. Spectroscopy on top of the cluster, away from the interface, shows Shiba states at finite positive and negative energies that are separated by a suppression of the tunnel conductance at zero energy (Fig. 1, B-C). A detailed analysis of the spatial profile of the ZBP in Fig. 4D demonstrates that it is strongest in magnitude on the superconducting 1D channel, precisely at the interface of the channel with the Fe cluster. The tail of the ZBP extends along the channel and as shown in Fig 4D can still be detected at about half of its peak value 2 nm away from the interface. This length scale is consistent with the theoretical expectation of the spatial decay of the MZM wavefunction in our system, when we crucially take into account the renormalization of the edge mode Fermi velocity caused by the proximity effect with the Nb substrate (see Sec. 9 of (*30*)) (*38-40*). We also observe the presence of such a ZBP localized at the hinge-cluster interface for clusters #1, #3 & #4 (see Sec. 8 and Fig. S11 of (*30*)). Overall, our experimental observation of a localized ZBP in these experiments is consistent with our model calculations described in Fig. 1, and its interpretation as a spectroscopic signature of an MZM.

Examining other clusters under varying experimental conditions, we obtain further observations that corroborate the interpretation of the ZBP as due to the presence of an MZM (see Sec. 10 of (*30*)). We find that the ZBP is absent when the application of a magnetic field suppresses superconductivity in our hybrid system, thereby excluding the Kondo effect as the origin of the ZBP (Fig. S13 of (*30*)). We also find that Fe clusters on top of B edges, where the edge modes are strongly hybridized with the bulk states, do not show ZBPs, suggesting that the presence of an isolated helical edge mode is required for the ZBP to appear (Fig. S14 of (*30*)). The data on these edges also demonstrate that the Fe clusters, even those as long as 80 Å, are not the source of the ZBP themselves (Fig. S15 of (*30*)). This is in contrast to other magnetic atomic chain systems on the surface of a superconductor (*4*), for which a non-trivial bulk topology of the chain itself can induce topological superconductivity and localize MZMs. Our theoretical modeling also predicts a second weakly localized MZM at the interface between an Fe cluster and a B edge owing to the hybridization of the topological edge state with the underlying bulk states on the B edge. Spectroscopic measurements near the B-edge find signatures of an LDOS enhancement near zero energy consistent with this prediction (see Sec. 9 of (*30*)). More importantly, although our observation of a ZBP has been reproduced for many superconducting A edge-Fe cluster interfaces (see Sec. 8 of (*30*)), we do find that a fraction of these systems does not show a ZBP. When comparing with the reconstructed magnetization vectors of the different Fe clusters shown in Fig. 3D, we observe a clear correlation between the observation of a ZBP and the type of cluster magnetization. Clusters #1-4 have a weak magnetization component perpendicular to the edge and show a ZBP, whereas clusters #5-6 have a sizable magnetization component perpendicular to the edge and do not show a ZBP. Such a behavior is consistent with our model calculations (Fig. 1E and Sec.1 of (*30*)), which show that a magnetization component perpendicular to the edge is detrimental to inducing the magnetic hybridization gap in the topological edge state necessary to localize an MZM.

**Spin polarization of MZMs**

Finally, we use energy-resolved spin-polarized spectroscopic measurements with the STM to demonstrate that we can distinguish the presence of an MZM from that of a trivial zero mode, which might form accidentally in our system. Previous studies have shown that trivial in-gap states appear with an energy asymmetric characteristic in the extracted spin-polarization $P(E)$ from spin-resolved STM spectroscopic measurements under constant current conditions *(6, 40)*. We have carried out such measurements on the Fe clusters, where we find the predicted behavior of P(E) for the Shiba states induced at low energies by these clusters (Fig. 5A). This asymmetric behavior of trivial states in P(E) results in their lacking spin contrast in this measurement at zero energy, and thereby can be used as a diagnostic tool to distinguish MZMs. As shown in Fig. 5B, the ZBP at the A-edge-Fe cluster interface shows a strong spin contrast at zero energy hence confirming that it does not arise from a trivial zero energy mode. Besides this important distinction between the MZM and trivial Shiba states, fundamentally in our platform the magnetization of these two different states are expected to point along different orientations. The spin orientation of the Shiba states is dictated by the orientation of the magnetic cluster, whereas the MZM occurring in the topological channel generically has a different spin orientation, which can be even perpendicular to the Shiba state. (see Sec. 11 of (*30*)). The differing sign of the spin polarization P(E) of the MZM and the positive energy Shiba state is consistent with this expectation and adds yet another unique feature of the MZM formed at the interface between superconductivity and magnetism on a topological edge mode. (see Sec. 11 of (*30*))

Looking ahead, our demonstration of MZMs within the topological edge states of Bi can be extended to Bi nanowires (*25*) and similar realization in other 2D or 3D higher order topological insulators by using magnetic clusters as a way of localizing MZMs in the presence of proximity-induced superconductivity. The Bi nanowires also provide a viable approach to create devices based on this platform. Clusters with weak magnetic anisotropy could be used to turn on and off the MZMs with in-plane magnetic fields reliably. If such fields are low enough, they might not disrupt the superconductivity of the edge but switch the magnetic state of the clusters. Employing spin-torque tunneling effects can present a viable alternative to switch the cluster magnetization, providing an approach to manipulate the presence of MZMs. In view of the recent progress in the field of 2D materials (*41-43*), we anticipate that this approach of localizing MZMs in topologically protected helical edge channels may also be realized using van-der-Waals heterostructures.

**Acknowledgments:** It is our pleasure to acknowledge initial theoretical support by Annika Johansson. **Funding:** This work has been primarily supported by Gordon and Betty Moore Foundation as part of EPiQS initiative (GBMF4530), ONR-N00014-17-1-2784, ONR-N00014-14-1-0330, NSF-MRSEC programs through the Princeton Center for Complex Materials DMR-142054, and the NSF-DMR-1608848. Additional support has come from the Alexander-von-Humboldt foundation through a Feodor-Lynen postdoctoral fellowship (BJ). Support for theory effort (BAB) has also come from DOE de- sc0016239, the NSF EAGER 1004957, Simons Investigator Grants, ARO MURI W911NF- 12-1-0461, the Packard Foundation, and the Schmidt Fund for Innovative Research. Theory effort has also been supported by National Natural Science Foundation of China under Project 11774317 (JL). **Author contributions:** BJ, YX, SJ and AY designed and conducted the experiment. BJ, YX and AY analyzed the data. BJ, JL and BAB performed the theoretical calculations. All authors contributed to the writing of the manuscript. **Competing interests:** The authors declare no competing interest. **Data and materials availability:** The experimental data and theory code of this study are available at (*44*).

**Supplementary Materials:**

Materials and Methods

Figures S1-S15

Tables S1 and S2

**Figures:**

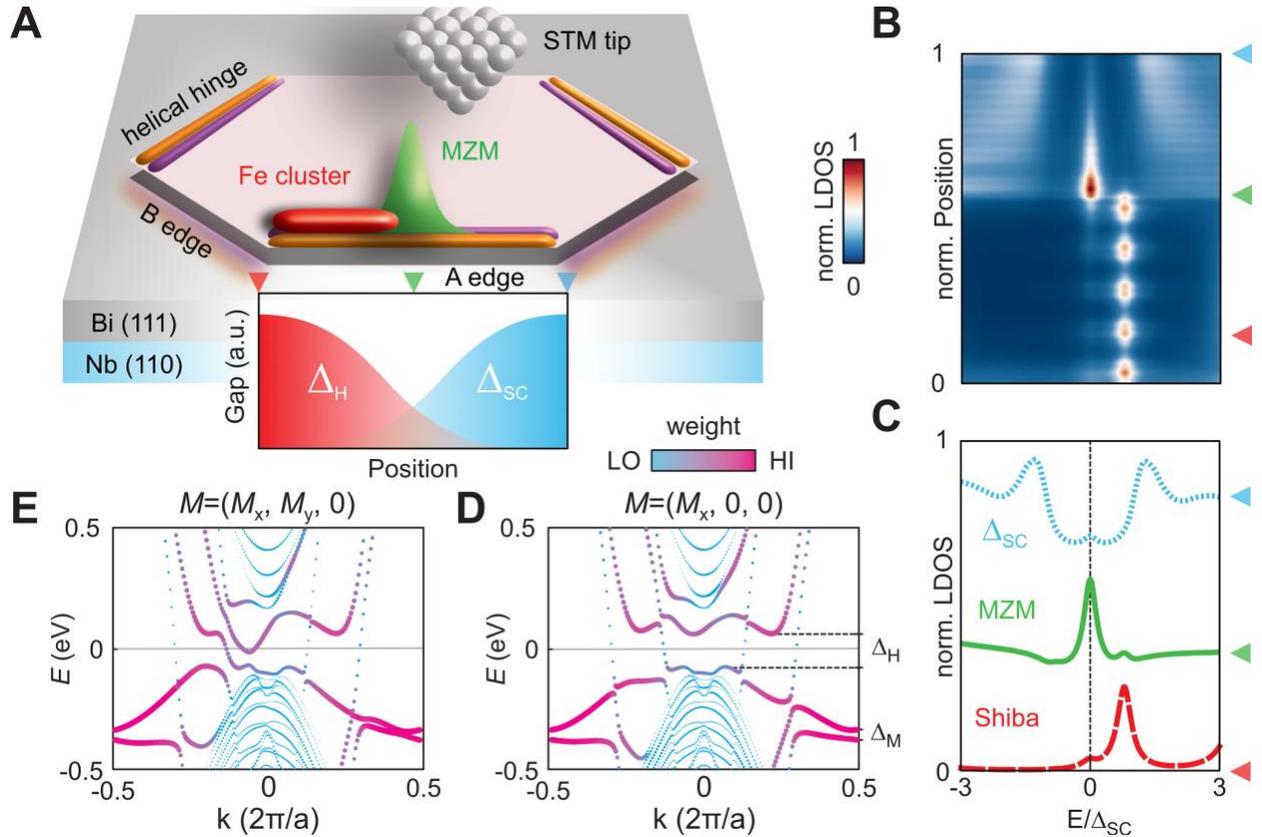

**Fig. 1. Topological Superconductivity and Majorana Zero Modes in the topological edge state of a Bi(111) bilayer (A)** Schematic representation of a hexagonal bismuth bilayer island sitting on the surface of a Bi(111) thin film and exhibiting topological helical states on every other edge. Topological superconductivity $\Delta_{SC}$ is induced into these helical states by superconducting proximity from the underlying Nb(110) substrate. Attaching a ferromagnetic cluster to the bilayer edge can open a magnetic hybridization gap $\Delta_H$. An MZM is localized at the mass domain wall realized at the cluster-helical edge state interface, and can be detected in STM experiments. **(B)** Spatially resolved low energy LDOS calculated from a tight binding model for the edge state cluster arrangement shown in (A) (see Sec. 1 of (*30*)). The LDOS is a spectroscopic line cut taken along the A edge in (A)**. (C)** Point-spectra extracted from the calculated spectroscopic line cut shown in (C) (positions indicated by the colored triangles). **(D)** and **(E)** Calculated band structure

along the Γ-M direction from a tight-binding model of a Bi(111) bilayer, for which the A edge is coupled to the spin-polarized d-bands of a ferromagnetic cluster resulting in a magnetic hybridization gap $\Delta_H$ and a Zeeman gap $\Delta_M$. For later reference, the Γ-M direction along the A edge is defined as the $\hat{x}$-direction. In **(D),** the cluster magnetization is parallel to the A edge $M = (M_x, 0, 0)$ and in **(E)** has an additional component of same amplitude perpendicular to the edge $M = (M_x, M_y, 0)$ (see Sec. 1 of (*30*)) The symbol size and color code encodes the wave function weight on the Bi(111) edge in contact with the cluster. The magnetic hybridization gap $\Delta_H$, spanning the entire Brillouin zone, as well as the Zeeman gap at the high-symmetry point $\Delta_M$ are indicated.

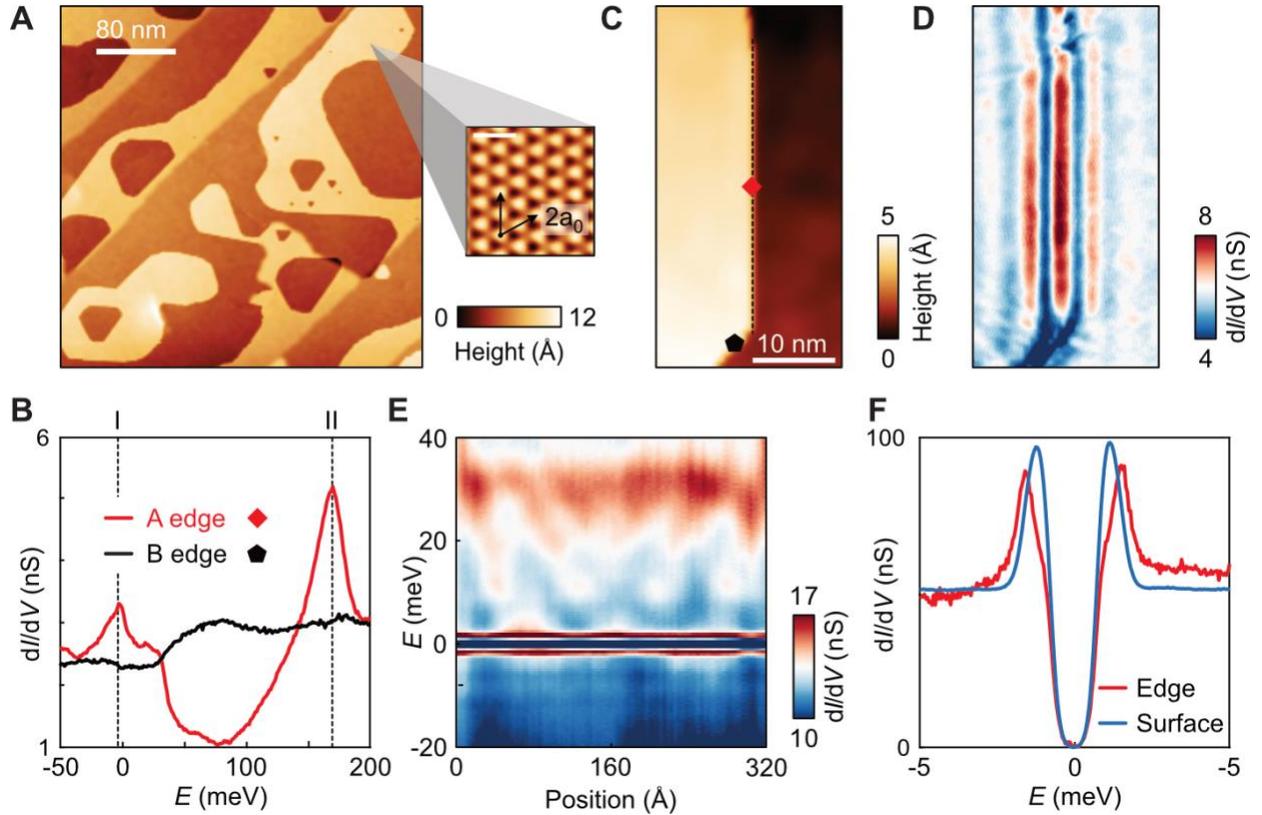

**Fig. 2. Structural and Electronic Properties of Bi(111) thin films.** (**A**) Large scale topography of the Bi(111) thin film on Nb(110). Inset: atomic lattice of the surface bilayer together with the lattice vector $a_0$ of the hexagonal lattice. (**B**) Point spectra measured on the A edge (red line) and B edge (black line), respectively at positions indicated in panel (C). The dashed lines labelled I and II denote energies at which van-Hove singularities in the edge state LDOS are observed ($V_{set}$ = 200 mV, $I_{set}$ = 2 nA, $V_{mod}$ = 1 mV). Spatial mapping at these features is demonstrated in (D) and Fig. S5 in (*30*). (**C**) Magnified topography of a typical A edge of a bilayer island, where the A edge is indicated by the red diamond and the B edge by the black pentagram. (**D**) Spatially resolved d$I$/d$V$ signal of the sample region shown in panel (C) ($V_{set}$ = 172 mV, $I_{set}$ = 2 nA, $V_{mod}$ = 0.3 mV). (**E**) One-dimensional spectroscopic line cut taken along the A-edge shown in panel (C) as indicated by the black dashed line in (C). The d$I$/d$V$ signal is plotted as a function of position and energy

($V_{set}$ = 40 mV, $I_{set}$ = 2 nA, $V_{mod}$ = 0.3 mV). **(F)** d$I$/d$V$ spectra measured in a small energy window around Fermi energy on the bilayer A edge (red) and on the bilayer surface (blue) ($V_{set}$ = 5 mV, $I_{set}$ = 1 nA, $V_{mod}$ = 40 µV).

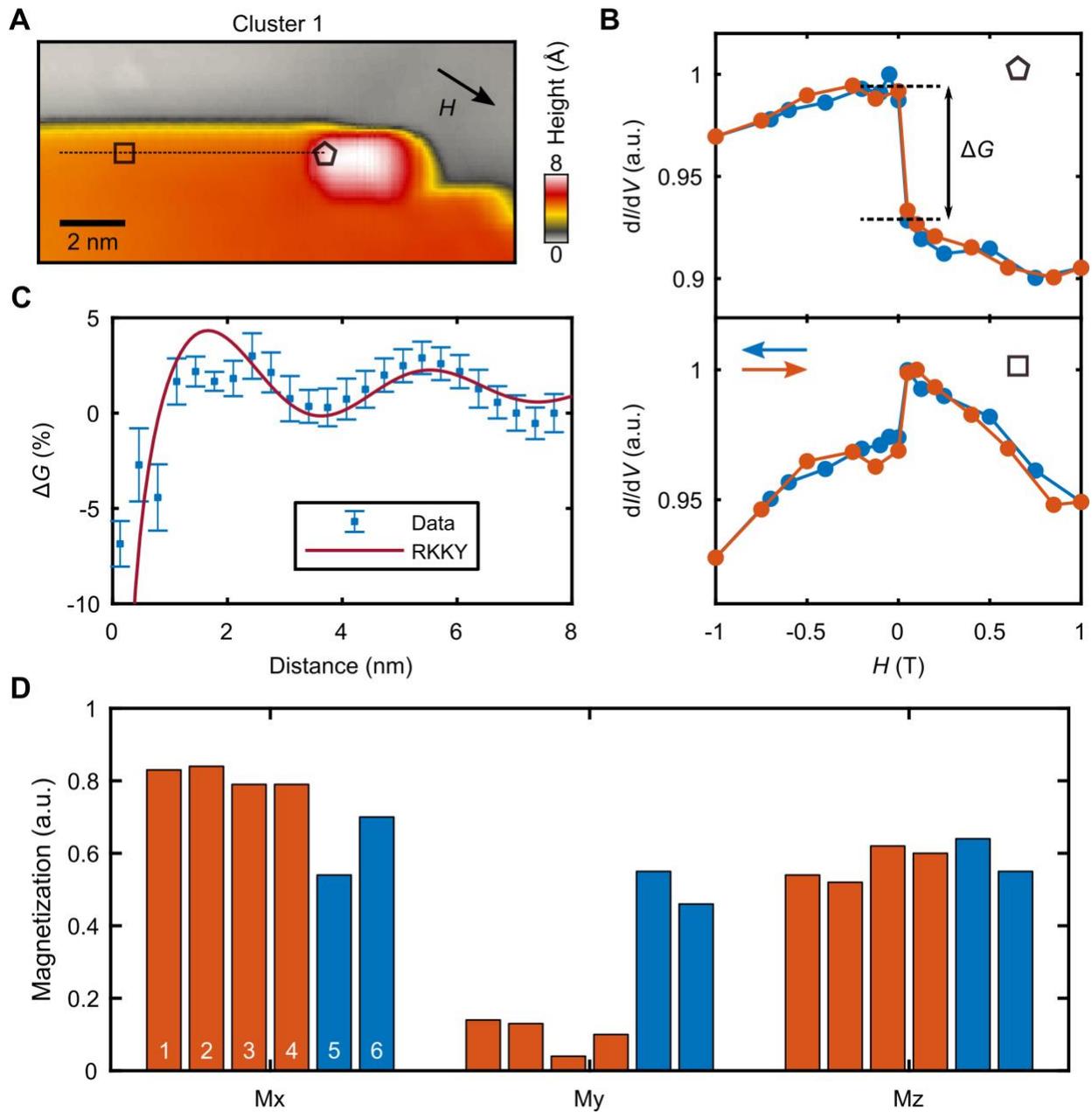

**Fig. 3. Spin polarized measurements on Fe cluster and Bi step edge.** (**A**) Topography of a Bi step edge decorated with Fe cluster #1. (**B**) Differential tunnel conductance $dI/dV$ measured on the Fe cluster (marked by pentagon) and on the bilayer step edge 6 nm away (marked by square) as a function of the applied magnetic field, the direction of which is indicated in panel (A). These $dI/dV$ spectra were recorded using a spin polarized Fe/Cr tip ($V_{set}$ = 10 mV, $I_{set}$ = 500 pA, $V_{mod}$ = 1 mV).

(**C**) Spatial dependence of the relative variation of conductance near zero field, $\Delta G$. The burgundy line is a fit using $J_1 - J_2\cos(2k_Fd)/(2k_Fd)$ (Fit parameters: $J_1 = 1.3$, $J_2=1.5$, $k_F=0.084$ Å$^{-1}$), with d being the distance away from the Fe cluster-hinge state interface along the dashed line in (A). (**D**) Bar plot of the $M_x$, $M_y$, and $M_z$ components of the normalized reconstructed magnetization vector for clusters #1-#6.

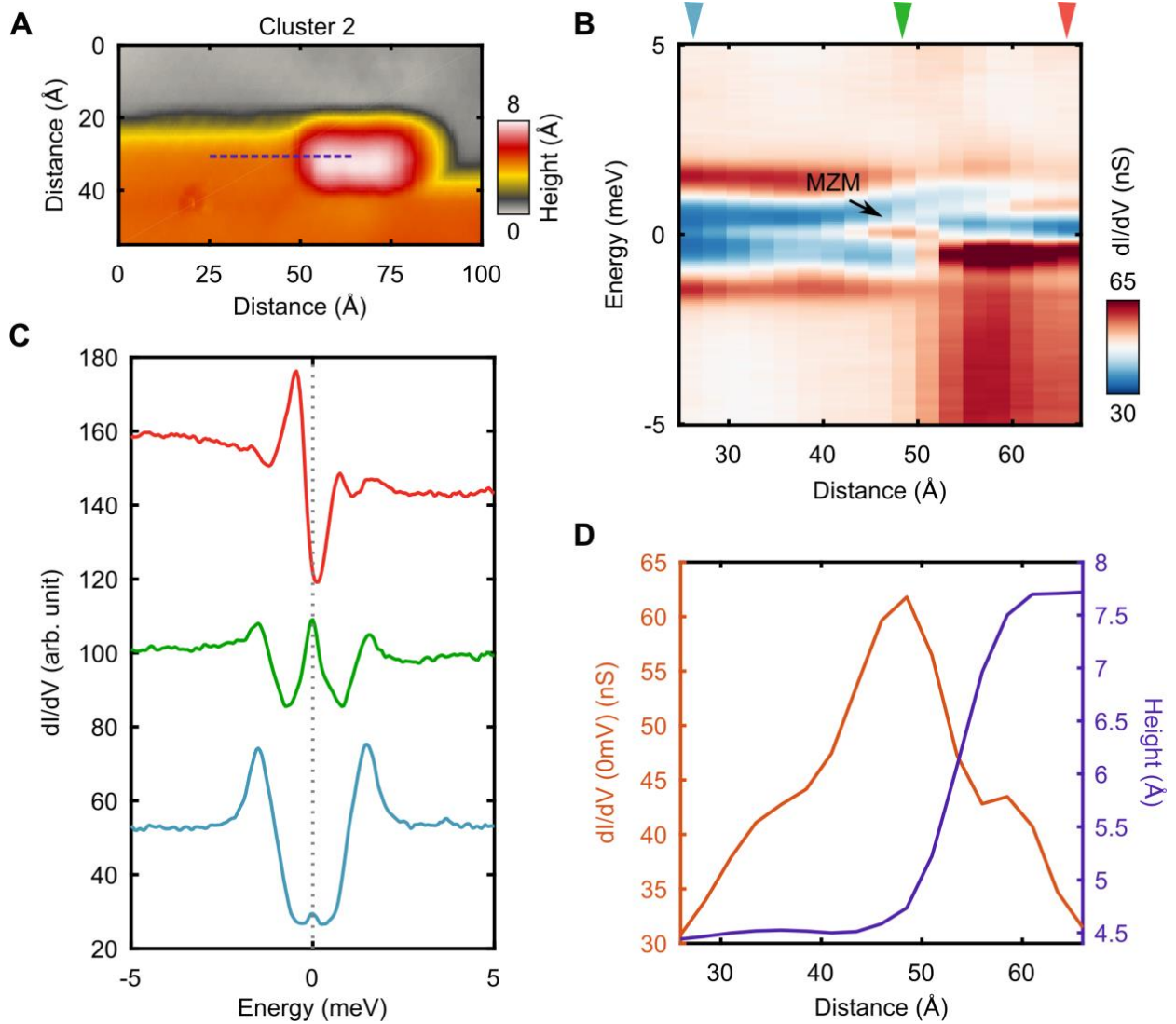

**Fig. 4. Localized ZBP at the interface.** (**A**) Topography of a Bi step edge decorated with Fe cluster #2. (**B**) Spectroscopic line cut taken along the purple dashed line shown in (A) ($V_{set}$ = 5 mV, $I_{set}$ = 1 nA, $V_{mod}$ = 40 μV). (**C**) Individual point spectra at locations indicated by triangular markers in (B). For clarity, the spectra are offset from each other by 45nS. (**D**) Simultaneously measured tunnel conductance d$I$/d$V$ at zero energy (red) and topographic height (purple) along the purple dashed line shown in (A).

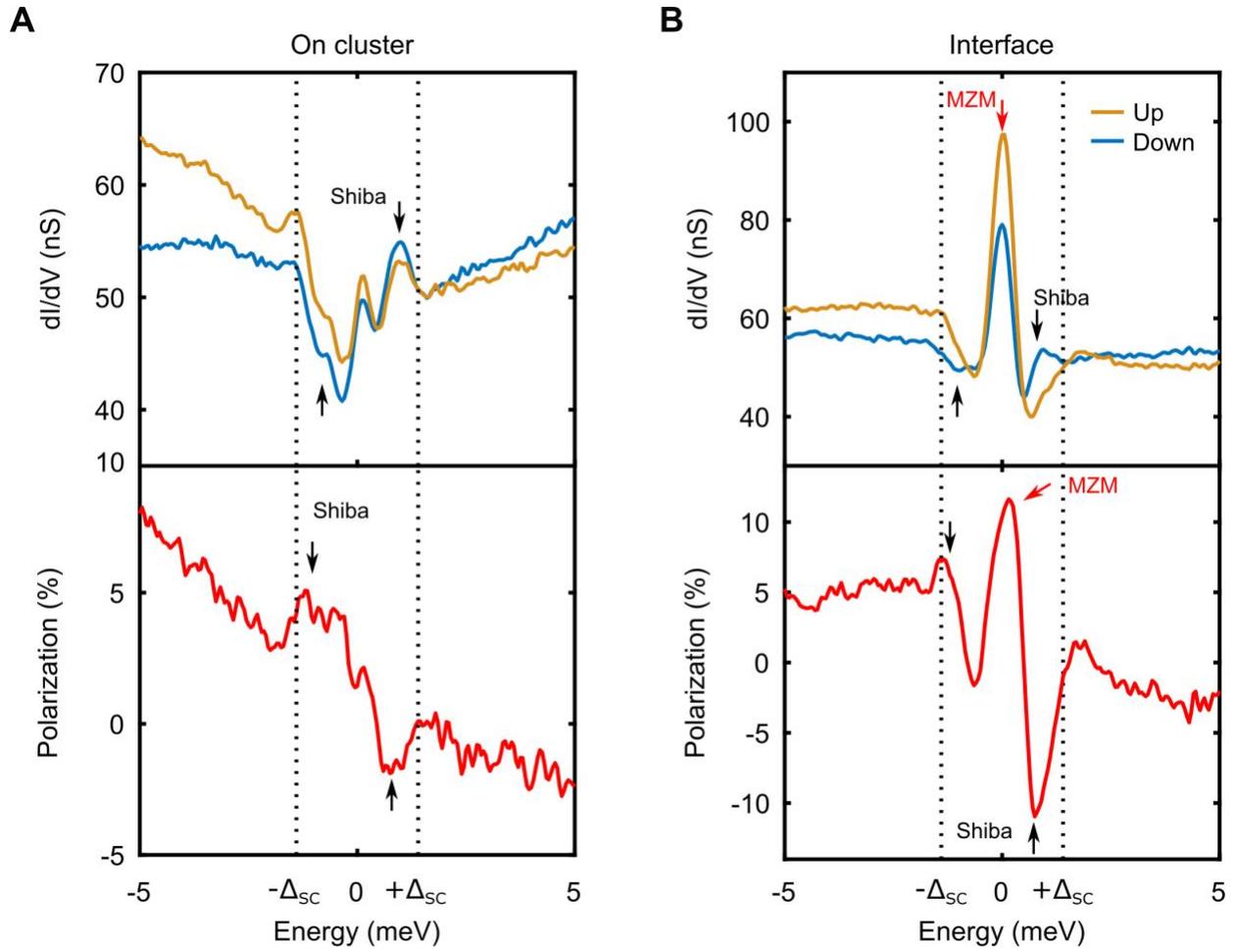

**Fig. 5. Spin polarization of MZM and Shiba states**. (**A** and **B**) d$I$/d$V$ spectra on the Fe cluster (A) and at the interface (B) as well as their corresponding spin polarizations ($V_{set}$=-5 mV, $I_{set}$=1 nA, $V_{mod}$=40 µV). Yellow and blue curves are taken with 'Up' and 'Down' polarized tips, respectively. Red arrows mark the zero-bias end state and black arrows mark the van-Hove singularity of the Shiba band.